\documentclass[
 superscriptaddress,
 amsmath,amssymb,
 twocolumn,
 aps,
 prd,
 10pt,
longbibliography
]{revtex4-2}

\usepackage{float}
\usepackage{graphicx,xcolor}
\usepackage{enumitem}
\usepackage{braket}
\usepackage[colorlinks=true]{hyperref}
\usepackage{amsfonts,amssymb}
\usepackage{amsmath, mathtools}
\usepackage{caption,subcaption}

\usepackage[normalem]{ulem}

\usepackage{float}
\usepackage{graphicx,xcolor}
\usepackage{tikz}
\usetikzlibrary{arrows.meta}

\usetikzlibrary{positioning}
\usepackage[edges]{forest}
\usepackage{enumitem}
\usepackage{braket}
\usepackage{hyperref}
\usepackage{amsfonts,amssymb}
\usepackage{amsmath, mathtools}
\usepackage{caption,subcaption}
\usepackage{comment}

\usepackage{amsthm}

\begin{document}
\title{Gravitational quantum speed limit 
}
\author{Nicola Pranzini}
\email{nicola.pranzini@helsinki.fi}
\address{QTF Centre of Excellence, Department of Physics, University of Helsinki, P.O. Box 43, FI-00014 Helsinki, Finland}
\address{InstituteQ - the Finnish Quantum Institute, Finland}


\author{Lorenzo Maccone}
\email{maccone@unipv.it}
\address{INFN Sez. Pavia, via Bassi 6, I-27100 Pavia, Italy}
\address{Dip. Fisica, University of Pavia, via Bassi 6, I-27100 Pavia, Italy}

\begin{abstract}
While playing an important role in the foundations of quantum theory, Quantum Speed Limits (QSL) have no role in discussions about the search for quantum gravity. We fill this gap by analysing what QSL arises when superposing spherically symmetric masses in canonical quantum gravity. By this procedure, we find that the quantum mechanical Mandelstam-Tamm and Margolus-Levitin bounds can be improved by superposing a spherically symmetric, static and asymptotically flat spacetime between states with different ADM energies and mass densities. We discuss the feasibility and significance of measuring times via these superpositions.
\end{abstract}

\maketitle

\section{Introduction}
Quantum Speed Limits (QSLs) are fundamental bounds on the minimum time required for a quantum system to evolve from some optimal state to an orthogonal one. As such, QSLs define how fast quantum processes can occur, constraining the speed at which quantum systems can manipulate information. Therefore, QSLs play a crucial role in quantum computing, quantum communication, and quantum control, setting theoretical limits on the performance and efficiency of quantum technologies. QSLs also have a foundational meaning regarding the nature of time and dynamics in quantum mechanics, highlighting the connection between time, energy, and the uncertainty principle relating them.

The two most widely known and used QSLs are the Mandelstam-Tamm (MT) and the Margolus-Levitin (ML) bounds~\cite{MandelshtamT45, MargolusL98}. Given a system with Hamiltonian $\hat{H}$, the first limit gives the minimal evolution time from a given state $\ket{\psi}$ to an orthogonal one as inversely proportional to the energy variance $\Delta_\psi E=[\braket{H^2}_\psi-\braket{H}^2_\psi]^{1/2}$ of the system~\cite{MandelshtamT45}. In contrast, the second relates the same quantity to the mean energy $E_\psi=\braket{H}_\psi-E_0$, where $E_0$ is the ground state energy~\cite{MargolusL98}. These can be combined into a unique QSL, which reads
\begin{equation}
    t_\perp\geq \max\left\lbrace\frac{\pi\hbar}{2E_\psi},\frac{\pi\hbar}{2\Delta_\psi E}\right\rbrace
    \label{e.QSL}
\end{equation}
\cite{GiovannettiEtAl03}. As it is clear, these bounds are based on the assumptions that a univocal time parameter is available and, of course, that non-relativistic quantum mechanics holds in general.

Both these assumptions are challenged by general relativity and Quantum Gravity. First, when discussing the time evolution of a localised system in relativity, it is essential to consider both where it occurs and the system's backreaction on the surrounding geometry, as time in relativity is dynamic and not universally defined. Moreover, when extending to QG, we encounter the so-called ``problem of time'', i.e. the discrepancy between general relativistic time - part of the dynamical spacetime geometry - and quantum mechanical time - an external parameter governing the system's evolution~\cite{Isham93, Kuchar94}. One way of solving the contradiction is quantising the spacetime itself, but this leads to the so-called frozen formalism: the quantisation procedure gives a Hamiltonian constraint, suggesting there is no explicit time evolution, making it difficult to reconcile the concept of time in quantum theory with general relativity~\cite{DeWitt67}. To address this issue, one possible approach is to extract a notion of time from within the system's dynamics, a process known as deparametrization, where time is identified with a physical degree of freedom rather than treated as an external parameter~\cite{Bergmann61}. Once the dynamics is unfrozen, it becomes possible to explore time evolution in QG; yet, the whole procedure is a complex issue that remains obscure in most aspects~\cite{MozotaFrauca23, CarlipH24}.

Inspired by their foundational meaning in QM, we explore how QSL change in a quantum gravitational setting. Specifically, this letter explores the generalization of the MT and ML bounds to a quantum gravitational scenario involving superpositions of a spherically symmetric massive body. In this way, we derive bounds on the speed of evolution of quantum spacetime, offering insights into the fundamental nature of low-energy quantum gravity (QG), and we find that the non-relativistic result is recovered when the massive system acting as a clock is observed from an asymptotic location or superposed between low mass-energy density states. Conversely, new QSLs emerge when the clock is superposed between states of vastly different mass and radius and observed from nearby. We will adopt the convention $c = \hbar = 1$ for the rest of this letter.


\section{Superposed spacetimes}
We start with a family of static, spherically symmetric and asymptotically flat four-dimensional spacetimes, labelled as $\mathcal{M}_{M,R}$ by their Komar masses
\begin{equation}
   M=\frac{1}{8\pi G} \int_{S}dSN^{ab}\nabla_a\xi_b
   \label{e.ADM_mass}
\end{equation}
(which, in this case, are equal to the ADM energies $E_{ADM}$) 
and radii $R$ for which the exterior Schwarzschild vacuum solution
\begin{equation}
    ds^2=-F_M(r)dt^2+F_M(r)^{-1}dr^2+r^2d\Omega^2
    \label{e.Schwarzschild}
\end{equation}
starts to be valid. In the above formulae, $\xi^a$ is a timelike Killing vector field normalised so that $V=(-\xi_a\xi^a)^{1/2}$ approaches one at infinity, $N^{ab}=2V^{-1}\xi^{[a}N^{b]}$ with $N^a$ being an outward pointing unit vector orthogonal to $\xi^a$ and normal to an asymptotic sphere $S$, and 
$F_M(r)=1-r_M/r$ with $r_M=2GM$ being the Schwarzschild radius proper of a body of mass $M$~\cite{Wald84}. When constructing our family of spacetimes, we suppose that all those configurations having $R\leq r_M$ describe a black hole and can be extended to cover zone II of a Kruskal diagram, while those having $R>r_M$ describe static non-black hole bodies and can be complemented by some interior metric
\begin{equation}
    ds^2=- A_M(r)dt^2 +B_M(r)^{-1}dr^2+r^2d\Omega^2
    \label{e.interior}
\end{equation}
satisfying the continuity conditions
\begin{equation}
    A_M(R)=B_M(R)=F_M(R)~.
\end{equation}
The specific forms of $A_M$ and $B_M$ can be obtained, for example, by imposing that the source of the gravitational field is made of an incompressible or isotropic fluid, fixing \eqref{e.interior} to give the interior Schwarzschild or Tolman--Oppenheimer--Volkoff metrics, respectively. 

Each spacetime in the above family can be foliated into spatial hypersurfaces with the topology of $\mathbb{R}^+\times S^2$, and hence be described by a collection of slices $\Sigma_t$ with three-metric 
\begin{equation}
    d\sigma^2=\Lambda^2(r,t)dr^2+R^2(r,t)d\Omega^2~,
\end{equation}
where $r\in[0,\infty)$, and $\Lambda(r,t)$ and $R(r,t)$ can be chosen to never vanish, so that the spatial metric is regular everywhere. Because our purpose is to describe the time evolution of a superposed body as seen from outside of its surface (be it a matter-vacuum interface or an event horizon), we need the ADM action for these spacetimes. Following the standard treatment of geometrodynamics and Kukcha\v{r}'s work on spherically symmetric solutions, for each spacetime we write the ADM action as
\begin{equation}
    S=\frac{1}{16\pi G}\int_{\mathbb{R}} dt \int_{\Sigma_t} d^3x \left[ N\sqrt{|h|}(K^{ab}K_{ab}-K^2+\mathcal{R})\right]
\end{equation}
where $N$ is the lapse function, and $h$, $K$ and $\mathcal{R}$ respectively are the 3-metric, the second fundamental form and the curvature scalar of $\Sigma_t$~\footnote{Thanks to spherical symmetry, $S$ can easily be integrated over the angular coordinates, thus giving an easy-to-manage two-dimensional ADM action.}. The action must be supplemented with a boundary term at infinity to ensure the variational principle is well-defined, produce correct equations of motion, and allow asymptotic energy to be a conserved quantity~\cite{ReggeT74}. Once supplemented with the boundary term, the action can be \textit{deparametrized} in terms of asymptotic proper time and used to derive the super-hamiltonian,  super-momentum, and mass conservation equations~\cite{Kuchar94}. Such deparametrization can be understood as unfreezing the time evolution in a constrained system by moving from an external time picture to an internal one. This method defines time in terms of the system's intrinsic variables (e.g. proper time of clocks at infinity) rather than by the external parameter labelling the spacetime leaves~\cite{Bergmann61}. For our purposes, it is important to note that the mass conservation equation 
\begin{equation}
    M(t)=M
    \label{e.mass_conservation}
\end{equation}
is valid in all foliation, asymptotic Killing, and proper times. Hence, because $M$ is a conserved quantity, it can serve as a physical observable that tracks time evolution classically~\cite{Greenberger10}; even when the Hamiltonian constraint is satisfied within the physical sector of the phase space, the variable canonically conjugated to the mass can effectively replace the external time established by the foliation, providing an internal physical time accessible to local observers in the asymptotic region of the spacetime. Thus, we can use Eq.~\eqref{e.mass_conservation} in both external and internal time to promote mass to a phase-space parameter spanning the family of spacetimes at fixed $R$ and later canonically quantise it to the mass operator $\hat{{\bf m}}$, thus achieving non-trivial time evolution in the canonically quantised picture.

Before moving to the quantisation of these spacetimes, let us notice that one can describe dynamics by several times: the foliation time $t$, the asymptotic proper time $\tau_\infty$, and the Killing time $T$. Although mathematically distinct and different in their meaning, all these times can be set to coincide asymptotically, thus giving the same physical dynamics at infinity~\cite{Kuchar94}. In this work, we want to describe time in yet another way: the proper time $\tau_i$ elapsed for an observer at fixed radial location $r_i$. Specifically, the latter can be chosen to be the object's surface, a choice relevant because it is the most appropriate way to describe measurements and experiments performed locally on the body and, hence, is better suited for a local quantum description of the problem. Because both $\tau_\infty$ and $\tau_i$ anchor at trajectories with constant radius outside the body, the latter is obtained from the former by
\begin{equation}
    \tau_i=\tau_\infty\sqrt{F_M(r_i)}~.
    \label{e.proper_times}
\end{equation}

We are now ready to read the above family of spacetimes as describing a single quantum object, which we can prepare in superpositions of configurations. The phase space of one single spacetime at fixed $M$ and $R$ can be extended to include other static, spherically symmetric and asymptotically flat spacetimes by promoting the mass and surface radius to phase space variables. Once this extended phase space is constructed, we quantise gravity by the so-called canonical quantisation scheme, performed for primordial black holes by K. V. Kucha\v{r} in Ref.~\cite{Kuchar94}. Amongst other phase space functional variables, the resulting quantum states are labelled by the couples $(M,R)$ and henceforth denoted by $\ket{\Psi_{M,R}}$. Constructing a well-defined Hilbert space hosting these states and the related inner product is still an open problem and an active area of research~\cite{AlonsoMonsalveElAl24, HeldM24}. Yet, assuming that the inner product can be defined so that the distinct configurations $\ket{\Psi_{M,R}}$ are classically distinguishable, i.e.
\begin{equation}
    \braket{\Psi_{M,R}|\Psi_{M',R'}}=\delta(M-M')\delta(R-R')~,
\end{equation}
we can construct spacetime superpositions as
\begin{equation}
    \ket{\Psi}=\int_0^\infty d\mu\int_{2G\mu}^\infty d\rho\, \gamma(\mu,\rho)\ket{\Psi_{\mu;\rho}}~,
    \label{e.superposition}
\end{equation}
with $p(\mu,\rho)=||\gamma(\mu,\rho)||^2$ the probability of finding the spacetime in the configuration $\ket{\Psi_{\mu,\rho}}$ and, consistently, $\iint
p(\mu,\rho)=1$~\cite{Halliwell90, ChenEtAl23}. 
Although the epistemic meaning of this probability is generally ambiguous due to the absence of a well-defined Hilbert space and inner product (as discussed above), we can, for the present discussion, invoke asymptotic flatness to interpret probability in the conventional sense using the Born rule~\cite{Kuchar91, Isham93, Kuchar11}. Finally, we remark that while the existence of superpositions of masses contrasts with Bargmann's superselection sector~\cite{Bargmann54}, it is now widely believed that their existence is required for a quantum relativistic theory to be internally consistent: Bargmann's superselection is not fundamental and should only be imposed in the non-relativistic regime~\cite{Greenberger01, Greenberger10}.

\section{Gravitational QSL} 
We are now ready to discuss the QSL one gets when measuring time by a quantum source of gravitational field. As discussed in the introduction, any time measurement in non-relativistic quantum mechanics is limited in its accuracy by some QSL; in general, this is a combination of the Mandelstam–Tamm (MT) and Margolus–Levitin (ML) bounds~\cite{GiovannettiEtAl03}. As it is clear from \eqref{e.QSL}, both bounds are expressed in terms of properties of the given state $\ket{\psi}$ and Hamiltonian $\hat{H}$, namely in terms of $\braket{\psi|\hat{H}^2|\psi}$ and $\braket{\psi|\hat{H}|\psi}$ for the first, and $\braket{\psi|\hat{H}|\psi}$ and $E_0$ for the second. Yet, the above quantization procedure for spacetimes forces all the physical states to satisfy the Wheeler-DeWitt constraint equation $\hat{H}\ket{\psi}=0$, hence depriving both $\Delta E_\psi$ and $\delta E_\psi$ of their meaning in this context. Consequently, a generalization of the above limits in QG must be performed in terms of some other observable that generates the system's evolution after deparametrization; $\hat{{\bf m}}$ provides such an operator. Yet, these bounds also have a direct operational interpretation: they describe the minimum time required to evolve from a given (energy superposition) state
\begin{equation}
    \ket{\psi}=\frac{1}{\sqrt{2}}\left(\ket{E_0}+\ket{E_1}\right)
\end{equation}
to an orthogonal one, since this state is optimal for attaining the QSL. For example, in the case of the MT limit, this results in
\begin{equation}
    t^\perp_{\rm ML}\geq\frac{\pi
    }{2|\delta E|}~,
    \label{e.ML}
\end{equation}
where $\delta E=(E_1-E_0)/2$ is the expectation value of the energy of $\ket{\psi}$~\cite{MandelshtamT45, MandelstamT91}. Following this lead, here we propose a gravitational version of the ML bound that, even if different in the type of eigenstates it superposes and dynamical generator it uses, retains the physical meaning of the above operational interpretation. 
As it will be clear, a similar approach can also yield a generalization of the MT limit.

Suppose we now measure times by a spherical source of gravitational field initially prepared in a uniform superposition of two mass and radius eigenstates. To this end, we place an observer in spacetime and ask them to wait for one unit of their proper time before comparing the system's state (i.e. the spacetime's state) with the initial one. First, consider an observer placed asymptotically to the gravitational source's location. Our previous construction gives the associated spacetime in the superposition
\begin{equation}
    \ket{\Psi(0)}=\frac{1}{\sqrt{2}}\left(\ket{\Psi_{M_0,R_0}}+\ket{\Psi_{M_1,R_1}}\right)~.
    \label{e.energy_superposition}
\end{equation}
Following \cite{Kuchar94}, in each branch the system evolves by the asymptotic propagator obtained from unparametrized canonical action as
\begin{equation}
    \ket{\Psi(\tau_\infty)}=\hat{U}_{\tau_\infty}\ket{\Psi(0)}=e^{-i\hat{{\bf m}}\tau_\infty
    }\ket{\Psi(0)}~,
\end{equation}
hence giving the implicit relation 
\begin{equation}
    \braket{\Psi(0)|\Psi(\tau_\infty^{\perp})}=0
\end{equation}
for the ML limit at asymptotic proper time $\tau_\infty^\perp$. By simple algebra, the above results in the gravitational version of the ML limit
\begin{equation}
    \tau_\infty^{\perp}\geq\frac{\pi
    }{2|\delta E_{\rm ADM}|}
    \label{e.asymptoticGSL}
\end{equation}
with $\delta E_{\rm ADM}=(M_1-M_0)/2$ the average energy. Therefore, an asymptotic observer can use a source of gravitational field as a quantum clock whose quantum speed limit is given by the ADM energies as in standard QM. It is important to remark here that, while the Wheeler-DeWitt constraint is satisfied by the above states, these still evolve by a Schr\"odinger equation expressed in terms of the mass operator $\hat{{\bf m}}$ and asymptotic proper time $\tau_\infty$. In practice, the mass operator is used to promote the external time, established by the foliation, to an internal physical time, seen by local observers in the asymptotic region of the spacetime.

Having established the QSL obtained by an asymptotic observer, we want to address what bound is seen from nearby the source. When placing the observer closer to the gravitating body, their position $r$ gets entangled with the geometric degree of freedom. Yet, because they have to wait for one click before checking orthogonality between states, the observer's proper time degree of freedom factors no matter what location they are placed at. In this case, different branches will experience different time evolution operators depending on the proper time elapsed for the observer in the branch. As a result, and recalling Eq.~\eqref{e.proper_times}, the time evolution operator local to the body is
\begin{equation}
    \hat{U}_\tau=\exp(-i\hat{{\bf m}}\tau
    )
    =\exp\left(-i\hat{{\bf m}}\tau_\infty\sqrt{F_M(r)}\right)~,
\end{equation}
i.e. it acquires an additional non-linear dependence on the ADM mass and observer's location degrees of freedom. When the above time evolution is applied to the superposed state of the source of gravitational field, one gets
\begin{equation}
    \ket{\Psi(\tau)}=\frac{1}{\sqrt{2}}\left(\hat{U}_{\tau}\ket{\Psi_{M_0,R_0}}+\hat{U}_{\tau}\ket{\Psi_{M_1,R_1}}\right)~,
\end{equation}
and hence
\begin{equation}
    \tau^\perp\geq\frac{\pi
    }{\left\lvert M_1\sqrt{F_{M_1}(r_1)}-M_0\sqrt{F_{M_0}(r_0)}\right\rvert}=\frac{\pi
    }{|2\delta E_{\rm ADM}-
    \mathcal{A}|}~,
    \label{e.GSL}
\end{equation}
where 
\begin{equation}
    \mathcal{A}=M_0\left[1-\sqrt{F_{M_0}\left(r_0\right)}\right]-M_1\left[1-\sqrt{F_{M_1}\left(r_1\right)}\right]
\end{equation}
represents the advantage or disadvantage of our gravitational ML bound over its non-gravitational counterpart. In the above expression, $r_i$ represents the observer's location in the $i$-th spacetime configuration appearing in the superposition. Because the denominator in the first line of Eq.~\eqref{e.GSL} is the difference between the effective gravitational mass as observed from a finite radius $r_i$ in each branch, $\mathcal{A}$ should be read as the deviation from the asymptotic value of the mass-energy contained in each branch as perceived by a local observer. Notice that, in the asymptotic limit, the observer's location gets disentangled from geometry and both coefficients $F_{M_i}(r_i)$ equal one, thus recovering Eq.~\eqref{e.asymptoticGSL}.

While in this section we focused on the Margolus-Levitin part of the QSL, identical considerations also hold for the Mandelstam-Tamm, since the state \eqref{e.energy_superposition} has energy standard deviation equal to the average of the energy of the superposed states. Thus, our discussion can be generalized to the full expression of the QSL reported in Ref.~\cite{GiovannettiEtAl03}.

\section{Discussion}
The limit obtained above depends on the difference between the Komar masses and densities of the superposed states, for $r_i$ is bounded below by $R_i$. Hence, looking at how $\mathcal{A}$ varies in its possible range of values 
\begin{equation}
    [-\min\{ M_0, M_1\},\max\{ M_0, M_1\}]
    \label{e.interval}
\end{equation}
as a function of the mass and densities appearing in the superposition, we gain insight into the modification introduced by quantum gravity to the standard ML bound. For the sake of clarity, we note that by choosing $\mathcal{A}$ in the above interval we span the range
\begin{equation}
    [\pi
    /\max\{M_0,M_1\}
    ,\infty)
    \label{e.time_interval}
\end{equation}
of lower bounds obtained by the gravitational ML limit; values in this latter interval vary in the former non-linearly, so that while the lower bound of \eqref{e.time_interval} corresponds to that of \eqref{e.interval}, the upper one is found for $\mathcal{A}=M_1-M_0$, which falls somewhere within the range \eqref{e.interval}. In what follows, we always fix $r_i=R_i$, for the most relevant modifications to the usual ML bound are obtained in these extreme cases. Indeed, as $r_i$ grows larger than $R_i$ in each branch, one gets a result closer and closer to the usual ML bound \eqref{e.ML} for the ADM energy - i.e. the result we already found by Eq.~\eqref{e.asymptoticGSL}. 

To study the behaviour of the gravitational ML limit in various situations, we start by considering the case where $\mathcal{A}\simeq 0$, corresponding to the body having a low density in both branches; in this case, our gravitational speed limit becomes the standard ML bound. This result shows that the quantum mechanical result can be recovered as a low-density limit of QG~\footnote{Remember that, for similar reasons, this also holds for the asymptotic limit of GQ, see above}. Next, we consider the higher and lower bounds of \eqref{e.interval}, which correspond to the one branch hosting a black hole and the other a low-density object; in this case, the gravitational ML limit outperforms the usual one. In particular, if the less massive branch is hosting the black hole, then the gravitational ML lower bound is smaller than the usual one, regardless of $\delta E$; this is because our QSL arises from the phase difference acquired by the two branches due to their difference in both masses and evolution rates, which can improve if time does not flow (or flows much slower) in either one. Finally, somewhere in the space of parameters, we find the case where both branches host a black hole ($\mathcal{A}= 2\delta E
$), for which the time measurement precision becomes null ($\tau^\perp\to\infty$); this is because the local time on the surface of the object does not flow in either branch, hence rendering meaningless the comparison between the times elapsed in each component of the superposition. Since the quantum speed limit arises from the phase difference acquired during the time evolution of each branch and given that, in this case, neither branch evolves, the time required to reach an orthogonal state diverges. As it is clear, considering the gravitational effects of the mass-energy eigenstates can improve the QSL, hence enhancing the precision we can achieve in time measurements.

Next, let us discuss \textit{if it is reasonable} and \textit{what it means} to perform time measurements using a superposition of spherical masses. First, we must address whether the required initial mass-energy superposition of spacetimes can be prepared in practice. To this end, suppose at asymptotic time $t_0$ an observer sends towards the test body some physical system that, on impact or absorption, applies the unitary operation $\hat{V}$ generating a rotation from the mass-energy eigenstate $\ket{\Psi_{M,R}}$ to the superposed state \eqref{e.energy_superposition}. For example, this can be a photon in an energy superposition generating the required superposition~\cite{FooEtAl22}, while the density tuning may be obtained by radiation pressure. Assuming this is possible, sending photons towards the body can be seen as an operation - performed locally and far from the test body - that applies the required unitary $\hat{V}$ on the body's state. Next, we should comment on the possibility of using the time-evolved state to measure time locally far from the test body. In the ML procedure, one directly tests the body and compares its initial and final states; if these are orthogonal, one can see that a tick of time has elapsed. On the contrary, in deriving our gravitational version of the QSL, we would like to require that some far-away observer performs all operations without direct access to the body (for this may even be a black hole). To this end, suppose the observer has access to both $\hat{V}$ and $\hat{V}^\dagger$; they start their procedure by applying $\hat{V}$ as described above, then they let some proper time $\tau$ flow and apply $\hat{V}^\dagger$. If the elapsed time is optimal (i.e. is an integer multiple of the ML time $\tau^\perp$),  $\hat{U}_\tau\hat{V}\ket{\Psi_{M,R}}$ is orthogonal to $\hat{V}\ket{\Psi_{M,R}}$, which means that $\hat{V}^\dagger\hat{U}_\tau\hat{V}\ket{\Psi_{M,R}}$ is orthogonal to the initial non-superposed state $\ket{\Psi_{M,R}}$: assuming this is a classically distinguishable state, the observer can then measure times by comparing the initial state with that obtained by sending the physical system realising $\hat{V}^\dagger$ at the test body. The question seems solved, but remembering what $\hat{V}$ was may raise some problems: it may be that applying $\hat{V}^\dagger$ requires the use of negative energy photons, for $\hat{V}$ entailed sending some (positive energy) photons at the test body. While the existence of negative energies is required by the combination of relativistic and field-theoretic principles~\cite{EpsteinEtAl65}, a complete discussion of whether $\hat{V}$ and $\hat{V}^\dagger$ exist must be carefully addressed; the discussion of this point goes beyond the scope of this letter.

Second, we address the meaning of measuring time by this procedure. According to the above paragraph, time measurements are achieved by starting from a mass-energy (time-invariant) and density eigenstate, putting it into a superposition of different mass-energy-densities via $\hat{V}$, letting the two branches evolve, applying $\hat{V}^\dagger$, and comparing the result with the initial state. If the two are orthogonal, we know that at least a ``click'' of time has elapsed. Looking at the expression \eqref{e.GSL}, it becomes clear that, although the procedure used to derive it is always operationally meaningful, the resulting minimal time does not have a direct physical interpretation in all cases. However, if one branch hosts a black hole with the observer placed on the event horizon, we get
\begin{equation}
    \tau^\perp\geq \frac{\pi
    }{M
    \sqrt{F_M(r)}}~,
    \label{e.GSL_w_BH}
\end{equation}
where $M$ and $r$ denote the mass and radius of the source in the other branch. In this case, the gravitational speed limit describes the time elapsed in the latter branch (this time is either proper of an observer at fixed $r$ or of an asymptotic one, depending on whether we multiply the inequality by $\sqrt{F_M(r)}$). Therefore, the presented procedure can measure time intervals for a given massive non-black hole body as follows. Starting from the state $\ket{\Psi_{M,R}}$, we use some $\hat{V}$ to make the superposition $\ket{\Psi_{M,R}}+\ket{\Psi_{M,R'<R_s}}$ with $R_s=2GM$, i.e. to generate a superposition of the starting body and a black hole with equal mass. Next, we apply the above prescription and find Eq.~\eqref{e.GSL_w_BH}. This way, the procedure reads as a projective measurement on the mass-energy-density eigenstate basis, and the information obtained by the measurement describes the proper, asymptotic, or Killing time elapsed.

\section{Acknowledgements}
The authors thank S. Maniscalco, A. Nyk\"anen, and all participants at the  QSAIL23 conference held around Isola d'Elba for interesting comments, discussions, and the amazing strength of will required for simultaneously sailing and thinking about quantum physics. NP thanks Y.Osawa and K. Numajiri for comments and discussions, the University of Pavia for hospitality while working on this manuscript, and acknowledges financial support from the Magnus Ehrnrooth Foundation and the Academy of Finland via the Centre of Excellence program (Project No. 336810 and Project No. 336814) and from the Research Council of Finland through the Finnish Quantum Flagship project (358878, UH). LM acknowledges support from the PRIN MUR Project 2022RATBS4.

\bibliographystyle{ieeetr}
\bibliography{bib}

\begin{thebibliography}{10}

\bibitem{MandelshtamT45}
L.~I. Mandelshtam and I.~E. Tamm, ``The uncertainty relation between energy and time in nonrelativistic quantum mechanics,'' {\em J. Phys. (USSR)}, vol.~9, p.~249–254, 1945.

\bibitem{MargolusL98}
N.~Margolus and L.~B. Levitin, ``The maximum speed of dynamical evolution,'' {\em Physica D: Nonlinear Phenomena}, vol.~120, no.~1, pp.~188--195, 1998.
\newblock Proceedings of the Fourth Workshop on Physics and Consumption.

\bibitem{GiovannettiEtAl03}
V.~Giovannetti, S.~Lloyd, and L.~Maccone, ``{The quantum speed limit},'' in {\em Fluctuations and Noise in Photonics and Quantum Optics} (D.~Abbott, J.~H. Shapiro, and Y.~Yamamoto, eds.), vol.~5111, pp.~1 -- 6, International Society for Optics and Photonics, SPIE, 2003.

\bibitem{Isham93}
C.~J. Isham, {\em Canonical Quantum Gravity and the Problem of Time}, pp.~157--287.
\newblock Dordrecht: Springer Netherlands, 1993.

\bibitem{Kuchar94}
K.~V. Kucha\ifmmode~\check{r}\else \v{r}\fi{}, ``{Geometrodynamics of Schwarzschild black holes},'' {\em Phys. Rev. D}, vol.~50, pp.~3961--3981, Sep 1994.

\bibitem{DeWitt67}
B.~S. DeWitt, ``Quantum theory of gravity. i. the canonical theory,'' {\em Phys. Rev.}, vol.~160, pp.~1113--1148, Aug 1967.

\bibitem{Bergmann61}
P.~G. Bergmann, ``Observables in general relativity,'' {\em Rev. Mod. Phys.}, vol.~33, pp.~510--514, Oct 1961.

\bibitem{MozotaFrauca23}
{\'A}.~Mozota~Frauca, ``Reassessing the problem of time of quantum gravity,'' {\em General Relativity and Gravitation}, vol.~55, p.~21, Jan 2023.

\bibitem{CarlipH24}
S.~Carlip and W.~Hu, {\em Covariant Canonical Quantization and the Problem of Time}, pp.~127--143.
\newblock Cham: Springer Nature Switzerland, 2024.

\bibitem{Wald84}
R.~M. Wald, {\em {General Relativity}}.
\newblock Chicago, USA: Chicago Univ. Pr., 1984.

\bibitem{Note1}
Thanks to spherical symmetry, $S$ can easily be integrated over the angular coordinates, thus giving an easy-to-manage two-dimensional ADM action.

\bibitem{ReggeT74}
T.~Regge and C.~Teitelboim, ``Role of surface integrals in the hamiltonian formulation of general relativity,'' {\em Annals of Physics}, vol.~88, no.~1, pp.~286--318, 1974.

\bibitem{Greenberger10}
D.~M. Greenberger, ``Conceptual problems related to time and mass in quantum theory,'' 2010.

\bibitem{AlonsoMonsalveElAl24}
E.~Alonso-Monsalve, D.~Harlow, and P.~Jefferson, ``{Phase space of Jackiw-Teitelboim gravity with positive cosmological constant},'' 2024.

\bibitem{HeldM24}
J.~Held and H.~Maxfield, ``{The Hilbert space of de Sitter JT: a case study for canonical methods in quantum gravity},'' 2024.

\bibitem{Halliwell90}
J.~J. Halliwell, ``Introductory lectures on quantum cosmology,'' in {\em Quantum Cosmology and Baby Universes}, WORLD SCIENTIFIC, 1991.

\bibitem{ChenEtAl23}
L.-Q. Chen, F.~Giacomini, and C.~Rovelli, ``Quantum {S}tates of {F}ields for {Q}uantum {S}plit {S}ources,'' {\em {Quantum}}, vol.~7, p.~958, Mar. 2023.

\bibitem{Kuchar91}
K.~V. Kucha\ifmmode~\check{r}\else \v{r}\fi{}, ``The problem of time in canonical quantum gravity,'' in {\em Conceptual Problems of Quantum Gravity} (A.~Ashtekar and J.~Stachel, eds.), Birkhauser, 1991.

\bibitem{Kuchar11}
K.~V. Kucha\ifmmode~\check{r}\else \v{r}\fi{}, ``Time and interpreatations of quantum gravity,'' {\em International Journal of Modern Physics D}, vol.~20, no.~supp01, pp.~3--86, 2011.

\bibitem{Bargmann54}
V.~Bargmann, ``On unitary ray representations of continuous groups,'' {\em Annals of Mathematics}, vol.~59, no.~1, pp.~1--46, 1954.

\bibitem{Greenberger01}
D.~M. Greenberger, ``Inadequacy of the usual galilean transformation in quantum mechanics,'' {\em Phys. Rev. Lett.}, vol.~87, p.~100405, Aug 2001.

\bibitem{MandelstamT91}
L.~Mandelstam and I.~Tamm, {\em The Uncertainty Relation Between Energy and Time in Non-relativistic Quantum Mechanics}, pp.~115--123.
\newblock Berlin, Heidelberg: Springer Berlin Heidelberg, 1991.

\bibitem{Note2}
Remember that, for similar reasons, this also holds for the asymptotic limit of GQ, see above.

\bibitem{FooEtAl22}
J.~Foo, C.~S. Arabaci, M.~Zych, and R.~B. Mann, ``Quantum signatures of black hole mass superpositions,'' {\em Phys. Rev. Lett.}, vol.~129, p.~181301, Oct 2022.

\bibitem{EpsteinEtAl65}
H.~Epstein, V.~Glaser, and A.~Jaffe, ``Nonpositivity of the energy density in quantized field theories,'' {\em Il Nuovo Cimento (1955-1965)}, vol.~36, pp.~1016--1022, Apr 1965.

\end{thebibliography}

\end{document}